\documentclass[journal]{IEEEtran}
\usepackage{graphicx}
\usepackage{amsmath,array}
\usepackage{etoolbox}
\usepackage{subfigure}
\usepackage{caption}
\usepackage{float}

\ifCLASSINFOpdf
\else
\fi
\begin{document}
%
\title{Carrier-Based Modulation Schemes Based on Symmetric Switching Patterns for Three-Phase Three-Switch Rectifier}
%
%
%

\author{Janamejaya Channegowda,~\IEEEmembership{Member,~IEEE,}
        Najath Abdul Azeez,~\IEEEmembership{Member,~IEEE,}
        and Sheldon S. Williamson ~\IEEEmembership{Fellow,~IEEE}}
\maketitle

\begin{abstract}
Electric Vehicle (EV) chargers are the need of the hour as EVs have invaded the commercial automobile market. The Three-Phase Three-Switch (TPTS) converter has features which make it ideal to act as the charging station because of its high efficiency at peak power rating. Traditionally, control of the TPTS converter has been achieved by the Space Vector (SV) based modulation scheme whose implementation is burdensome due to the trigonometric calculations involved. This paper introduces two simplified carrier based modulation schemes, which greatly reduce the implementation difficulty, as they require minimum  mathematical computations. A quantitative comparison is made between SV scheme available in literature with the simplified scheme proposed here. The methodology followed to obtain the carrier waves is also provided. All the modulation schemes are validated with the help of a 1 kW hardware prototype.
\end{abstract}


\section{Introduction}\label{sec1}

\begin{figure}[hbpt!]
\centering
\includegraphics[scale=0.5]{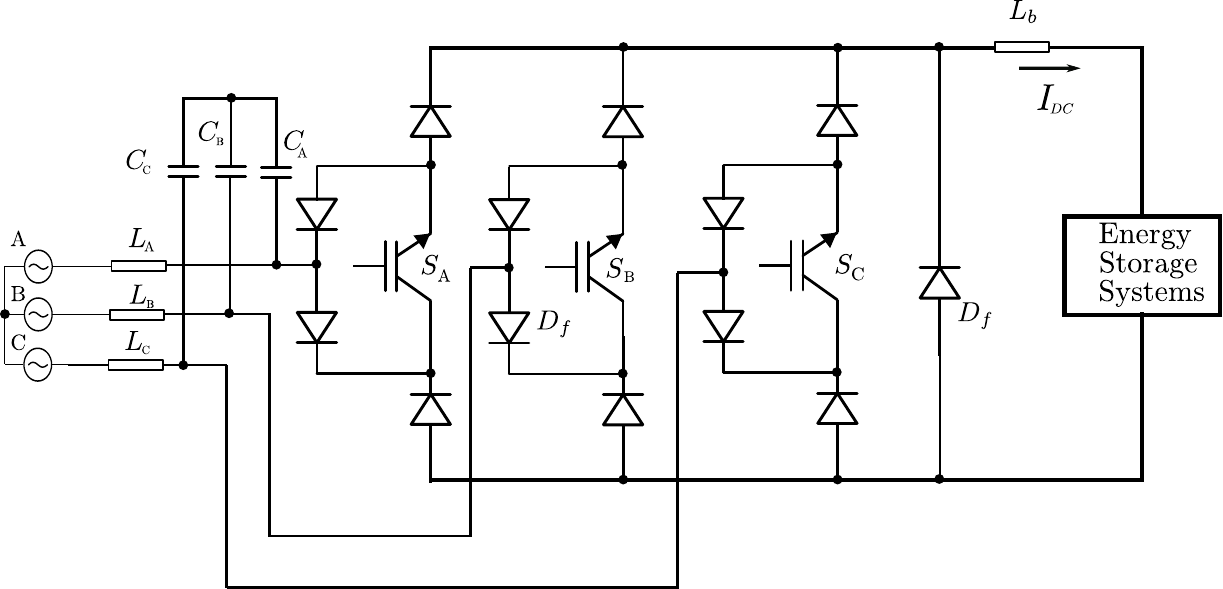}
\caption{Three Phase Three Switch front end rectifier}
\label{fig:tpts}
\end{figure}

Charging stations have sprung up in great numbers over the last decade \cite{evs}. However, in order to charge the EV battery in a short duration, the charging station should have a converter with high power rating \cite{ptk}. Traditionally, due to the power requirement, three phase converters have been employed in Fast Charging (FC) stations \cite{t1}. There is a need to reduce this two-staged converter topology to an efficient single-staged converter. Several topologies have been explored to achieve this feat \cite{jan}. TPTS converter is a strong  contender to be used in the FC station as it has features such as: a) Wide output voltage range; b) Low losses at higher power rating \cite{kodes}.  The TPTS converter as shown in Fig \ref{fig:tpts} has only one active switch in each leg, hence, greatly reducing the switching losses in the converter \cite{newwide}. Control of TPTS converter was through Space Vector (SV) based modulation scheme, this was predominantly used in all its analysis. Though SV modulation has several benefits, such as: i) Low switching loss; ii) Reduction in  current ripple of the DC link inductor \cite{k1}, implementation on digital systems is tough due to the number of trigonometric computations involved in calculating dwell times. There have been a few endeavors in the past to use carrier-based modulation schemes for the TPTS converter \cite{car1}. The focus of this paper is to simplify the modulation technique for the TPTS converter by employing various carrier based modulation schemes. The carrier waves are generated from symmetric switching sequences. The various carrier waves generated from switching patterns are described in Section II. Control algorithm along with details about the hardware setup are provided in Section III. Simulation and experimental results are presented in Section IV, followed by the conclusion in Section V. 

\begin{figure}[hbpt!]
\centering
		\includegraphics[scale=0.18]{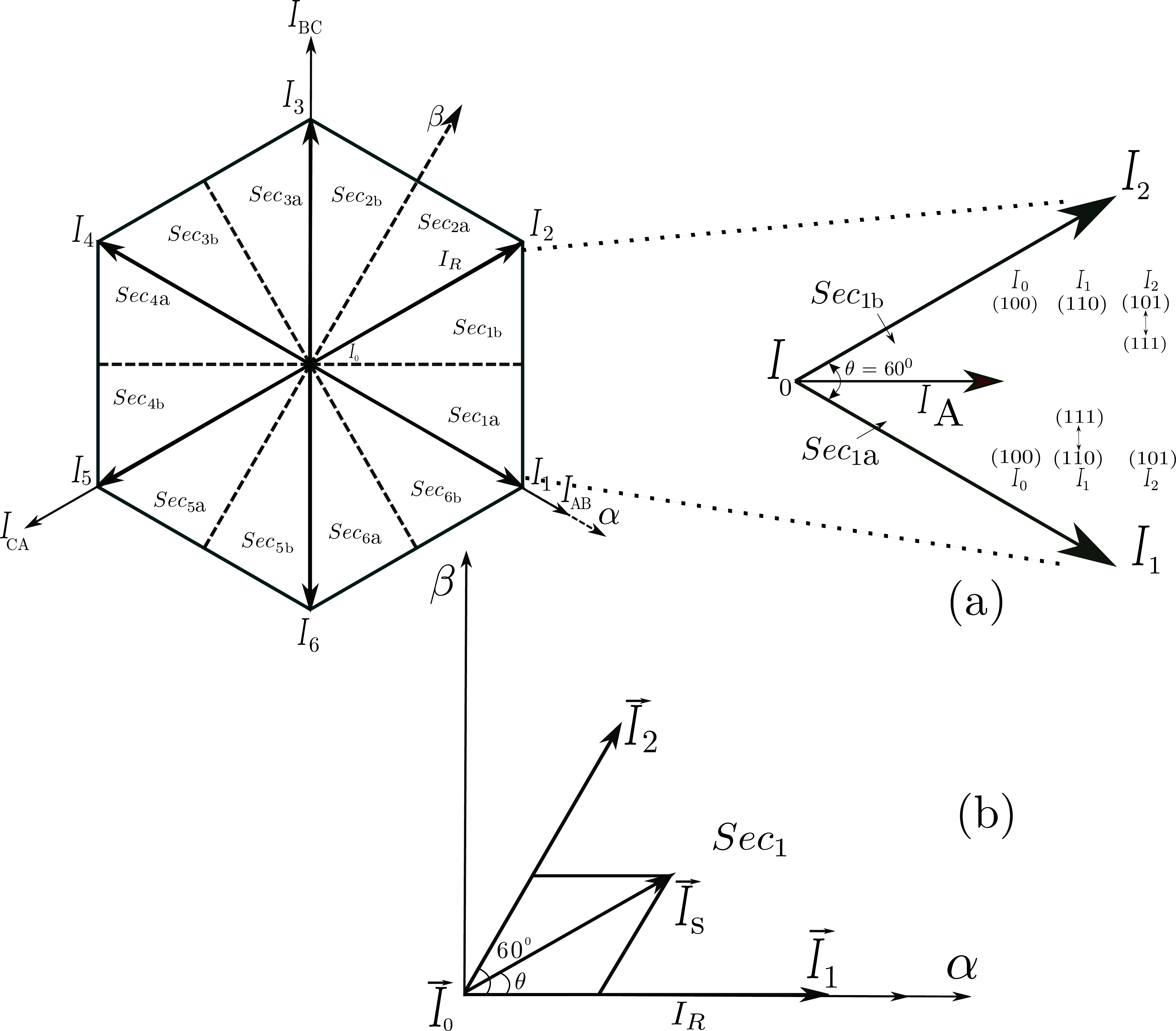}
	\caption{(a) Current space vector switching patterns, (b) Space vector decomposition in Sector 1.}
	\label{fig:two_sec_ia}
\end{figure}
 
 \begin{table*}[t] 
 \caption{Arrangement of active and zero states in subsector 1a and 1b for various switching patterns}
      \centering
\begin{tabular}{ >{\centering\arraybackslash}m{1in}  >{\centering\arraybackslash}m{2in}  >{\centering\arraybackslash}m{2in}}
\noalign{\smallskip}\hline\noalign{\smallskip}
\textbf{Pattern} &  \begin{tabular}{c} \textbf{Subsector 1a} \\$\pmb{T_0}$ \hspace{0.2cm} $\pmb{T_2}$ \hspace{0.25cm} $\pmb{T_1}$  \hspace{0.2cm} $\pmb{T_1}$ \hspace{0.2cm} $\pmb{T_2}$ \hspace{0.2cm} $\pmb{T_0}$\end{tabular} &
\begin{tabular}{c}\textbf{Subsector 1b} \\$\pmb{T_0}$ \hspace{0.2cm} $\pmb{T_1}$ \hspace{0.2cm} $\pmb{T_2}$ \hspace{0.25cm} $\pmb{T_2}$ \hspace{0.2cm} $\pmb{T_1}$ \hspace{0.2cm} $\pmb{T_0}$\end{tabular}
  \\ \hline
 
\noalign{\smallskip}\noalign{\smallskip}
I & 001 \hspace{0.1cm} 101 \hspace{0.1cm} 111 \hspace{0.1cm} 111 \hspace{0.1cm}   101 \hspace{0.1cm}   001  & 010 \hspace{0.1cm} 110 \hspace{0.1cm} 111 \hspace{0.1cm} 111 \hspace{0.1cm}   110 \hspace{0.1cm}   010   \\ \hline
\noalign{\smallskip}\noalign{\smallskip}
II & 000 \hspace{0.1cm} 101 \hspace{0.1cm} 111 \hspace{0.1cm} 111 \hspace{0.1cm}   101 \hspace{0.1cm}   000  & 000 \hspace{0.1cm} 110 \hspace{0.1cm} 111 \hspace{0.1cm} 111 \hspace{0.1cm}   110 \hspace{0.1cm}   000   \\ \hline
\end{tabular}
\centering 
       \label{tab:arrange vect}
         \end{table*}

\section{Carrier Based PWM for TPTS}
Carrier based modulation schemes have been used to control power converters due to their simplicity of implementation \cite{emil}. In conventional SV based modulation scheme for TPTS, placement of active and zero vectors within a switching time period coupled with the selection of redundant vectors, leads to multiple PWM schemes as shown in Fig. \ref{fig:two_sec_ia}a \cite{naj},\cite{janapec}. Rearrangement of active and zero vectors in SV modulation scheme for TPTS, leads to multiple switching patterns. A subset of these switching patterns can be used to generate carrier waves.

The TPTS converter has four zero vectors obtained from the following states, $s_j$ = 000, 001, 010, 100, which are common to all subsectors. The available active vectors are represented by the following states, 110, 101, 011 and 111. The TPTS converter has a redundant active vector got from the state 111 \cite{newwide}. In order to derive carrier waves from the SV modulation schemes for TPTS converter, following conditions have to be met: 

\begin{itemize}
\item In switching time period, $T_s$, none of the three phases should switch more than once
\item The redundant active state is placed in the center of the symmetric switching pattern and should always be employed
\item The other active state is placed beside the redundant state
\end{itemize}
 
It can also be observed that, in the switching sequences generated from these conditions, using zero states 100, 010, and 001 leads to one of the phases to be clamped. Various switching patterns in subsectors 1a and 1b, which conform to the conditions mentioned earlier are listed in Table \ref{tab:arrange vect}. Following sections provide details related to the carriers generated from these switching patterns.

\subsection{Switching Pattern I}
\begin{figure}[htbp!]
\centering
\includegraphics[scale=0.18]{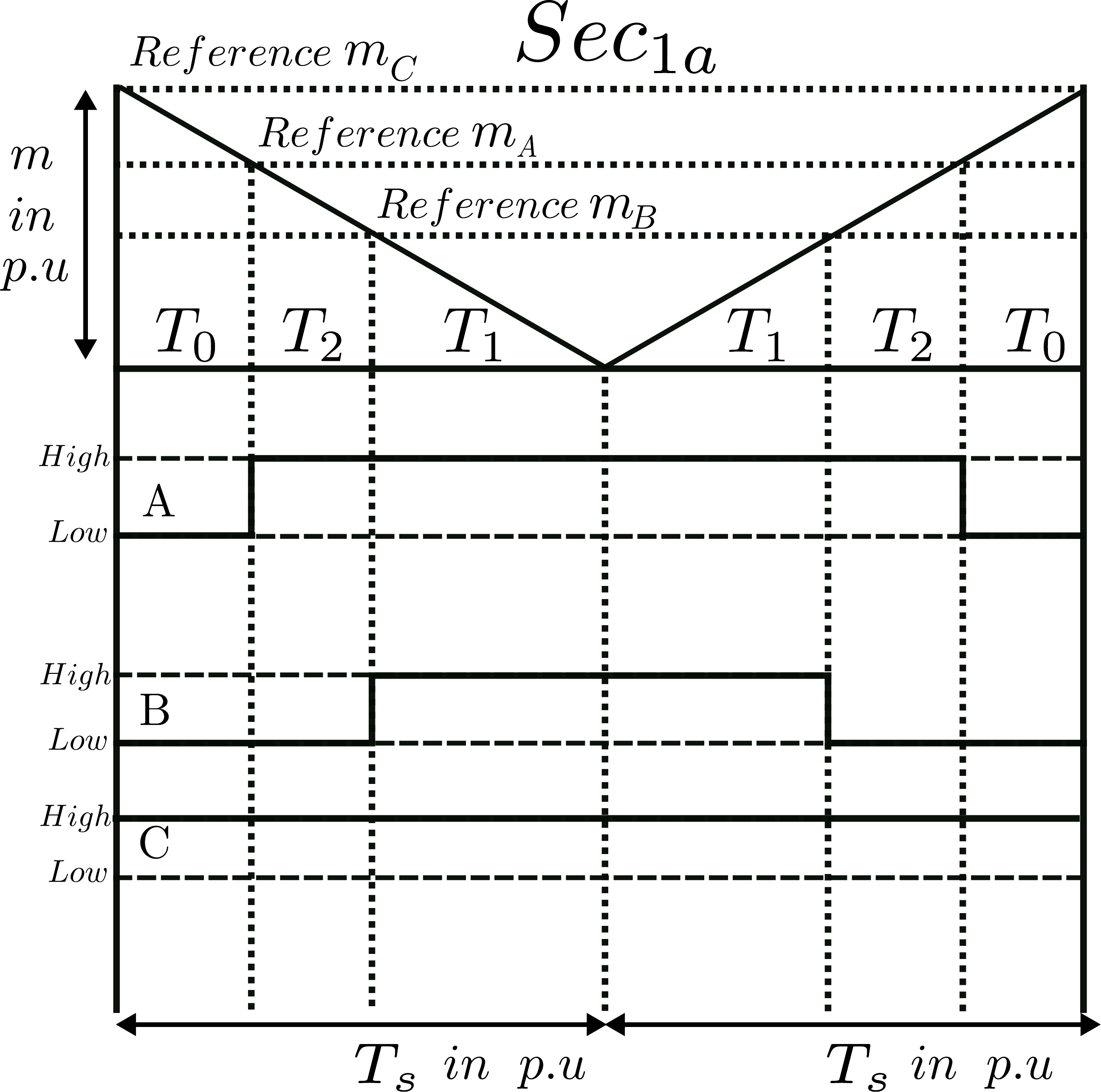}
\caption{Switching Pattern I in Subsectors 1a and 1b}
\label{fig:tri_drop_II_a}
\end{figure}

\begin{figure}[htbp!]
\centering
\includegraphics[scale=0.18]{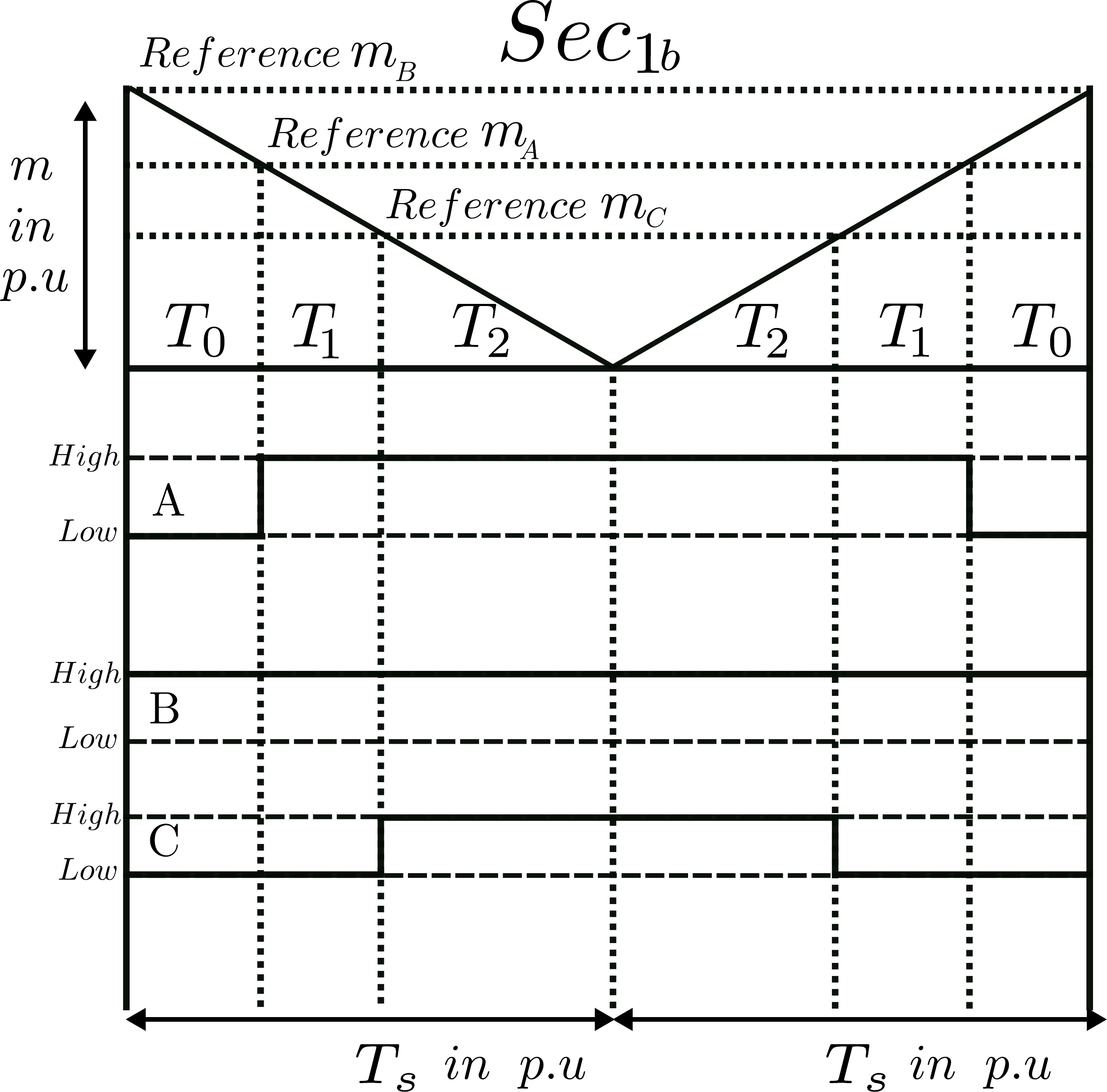}
\caption{Switching Pattern I in Subsectors 1a and 1b}
\label{fig:tri_drop_II_b}
\end{figure}

In subsector 1a, the current vectors $I_0$ (001), $I_2$ (101) and $I_1$ (111) are arranged in dwell times $T_0$, $T_2$ and $T_1$ as shown in Fig \ref{fig:tri_drop_II_a}. In subsector 1b, the current vectors  $I_0$ (010), $I_1$ (110), $I_2$ (111) are arranged in dwell times $T_0$, $T_1$ and $T_2$ as shown in Fig \ref{fig:tri_drop_II_b}. The on-times  $T_{on,A}$, $T_{on,B}$, $T_{on,C}$  for the three phase currents $I_A$, $I_B$ and $I_C$  for subsector 1a is given by (1) - (3)  and for subsector 1b is given by (4) - (6).
\begin{align}
T_{on,A} &= T_2+T_1 = \frac{T_s}{I_{DC}} (-I_C-I_B) = \frac{T_s}{I_{DC}} \cdot I_A \\
T_{on,B} &= T_1 = \frac{T_s}{I_{DC}} \cdot -I_B \\
T_{on,C} &= T_0+T_1+T_2 = T_s \\
T_{on,A} &= T_1+T_2 = \frac{T_s}{I_{DC}} (-I_B-I_C) = \frac{T_s}{I_{DC}} \cdot I_A \\
T_{on,B} &= T_0+T_1+T_2 = T_s \\
T_{on,C} &= T_2 = \frac{T_s}{I_{DC}} \cdot -I_C
\end{align}
Scrutinizing the on-times for this switching pattern, a generalized expression for the on-times $T_{on,maxAbs}$, $T_{on,midAbs}$ and $T_{on,minAbs}$ of $I_{maxAbs}$, $I_{midAbs}$ and $I_{minAbs}$ are given by (7) - (9), where $I_{maxAbs}$, $I_{midAbs}$ and $I_{minAbs}$ respectively represent the maximum, middle and minimum amplitudes of the absolute values of the three-phase reference currents. It has to be noted that in this pattern, the phase with the minimum absolute amplitude, in a particular sub-sector, is always clamped for $T_s$ duration.
\begin{align}
T_{on,maxAbs} &= I_{maxAbs} \cdot \frac{T_s}{I_{DC}} \\
T_{on,midAbs} &= I_{midAbs} \cdot \frac{T_s}{I_{DC}} \\
T_{on,minAbs} &= T_s 
\end{align}
 
\subsection{Switching Pattern II}
\begin{figure}[htbp!]
\centering
\includegraphics[scale=0.18]{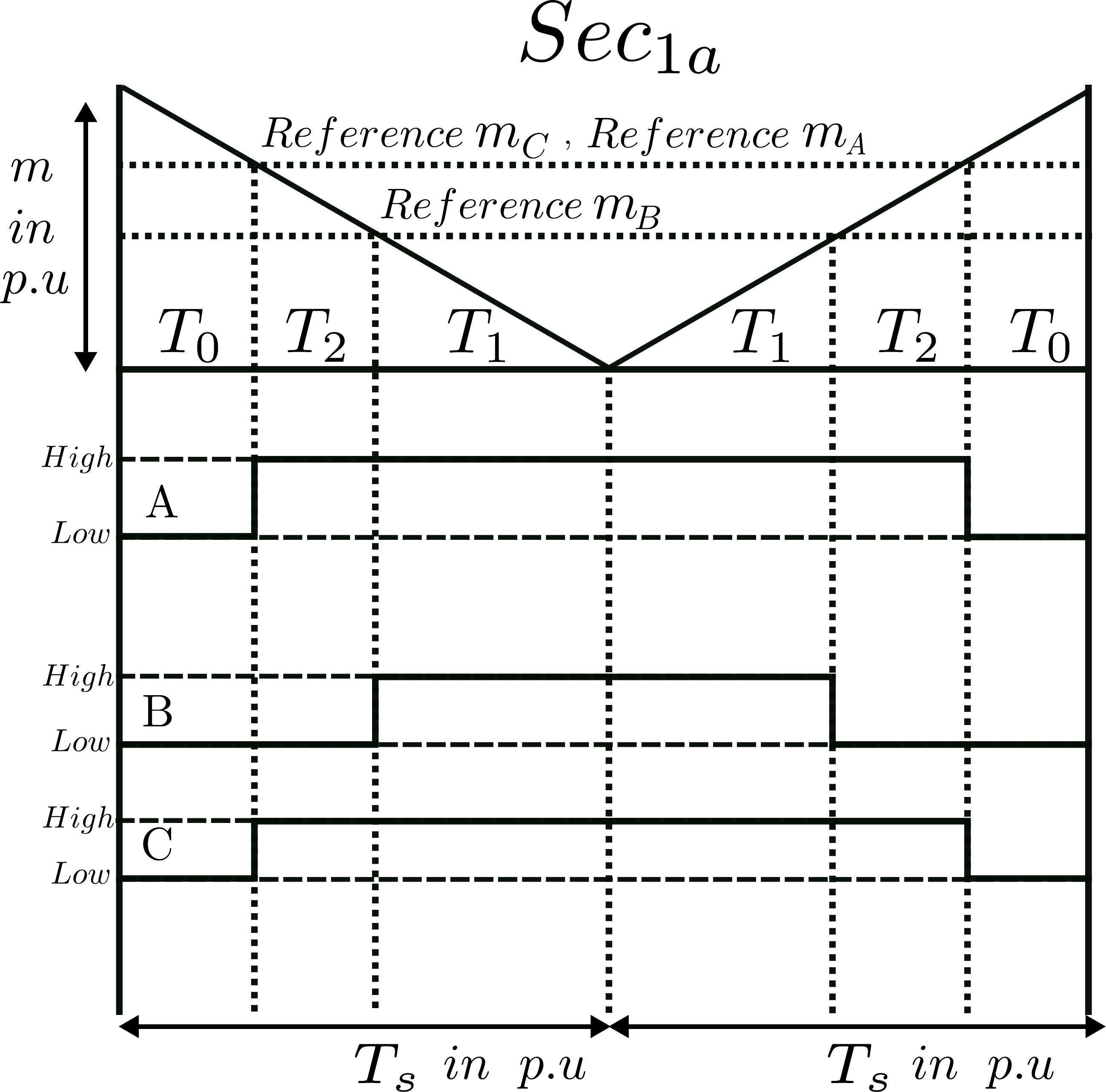}
\caption{Switching Pattern II in Subsectors 1a and 1b}
\label{fig:tri_drop_III_a}
\end{figure}

\begin{figure}[htbp!]
\centering
\includegraphics[scale=0.18]{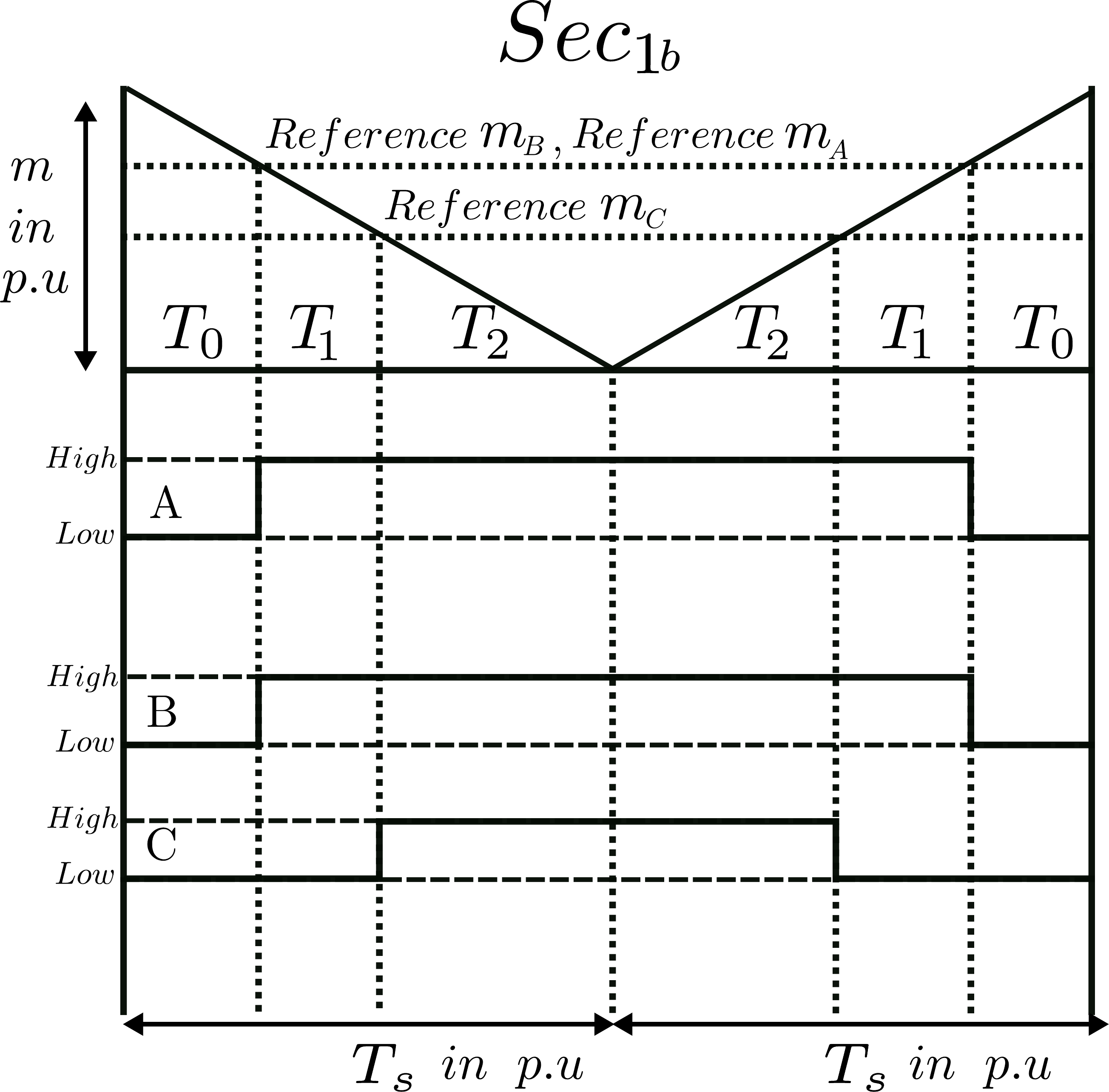}
\caption{Switching Pattern II in Subsectors 1a and 1b}
\label{fig:tri_drop_III_b}
\end{figure}

In subsector 1a, the current vectors $I_0$ (000), $I_2$ (101) and $I_1$ (111) are arranged in dwell times $T_0$, $T_2$ and $T_1$ as shown in Fig \ref{fig:tri_drop_III_a}. In subsector 1b, the current vectors  $I_0$ (000), $I_1$ (110), $I_2$ (111) are arranged in dwell times $T_0$, $T_1$ and $T_2$ as shown in Fig \ref{fig:tri_drop_III_b}. The on-times  $T_{on,A}$, $T_{on,B}$, $T_{on,C}$ for the three phase currents $I_A$, $I_B$ and $I_C$  for subsector 1a is given by (10) - (12) and for subsector 1b is given by (13) - (15).

\begin{align}
T_{on,A} &= T_2+T_1 = \frac{T_s}{I_{DC}} (-I_B-I_C) = \frac{T_s}{I_{DC}} \cdot I_A \\
T_{on,B} &= T_1 = \frac{T_s}{I_{DC}} \cdot -I_B \\
T_{on,C} &= T_2+T_1 =  \frac{T_s}{I_{DC}} (-I_B-I_C) =  \frac{T_s}{I_{DC}} \cdot I_A \\
T_{on,A} &= T_2+T_1 = \frac{T_s}{I_{DC}} (-I_B-I_C) = \frac{T_s}{I_{DC}} \cdot I_A \\
T_{on,B} &= T_2+T_1 = \frac{T_s}{I_{DC}} (-I_B-I_C) = \frac{T_s}{I_{DC}} \cdot I_A \\
T_{on,C} &= T_2 = \frac{T_s}{I_{DC}} \cdot -I_C
\end{align}

Inspecting the on-times we arrive at a generalized expression for the on-times $T_{on,maxAbs}$, $T_{on,midAbs}$ and $T_{on,minAbs}$ of $I_{maxAbs}$, $I_{midAbs}$ and $I_{minAbs}$ are given by (16) - (18), where $I_{maxAbs}$, $I_{midAbs}$ and $I_{minAbs}$ respectively represent the maximum, middle and minimum amplitudes of the absolute values of the three-phase reference currents. This pattern differs from pattern I as it can be observed that none of the phases are clamped in any subsector. All the ON-times of the two carrier waves can be compared as seen in Table \ref{tab:discussion}.
 
\begin{align}
T_{on,maxAbs} &= I_{maxAbs} \cdot \frac{T_s}{I_{DC}} \\
T_{on,midAbs} &= I_{midAbs} \cdot \frac{T_s}{I_{DC}} \\
T_{on,minAbs} &= I_{maxAbs} \cdot \frac{T_s}{I_{DC}} 
\end{align}
 
%
%

\begin{table}[htbp!] 
\caption{Comparison of ON-Times of Switching Patterns}
      \centering
       \begin{tabular}{ >{\centering\arraybackslash}m{0.6in}  >{\centering\arraybackslash}m{0.7in}  >{\centering\arraybackslash}m{0.7in}
>{\centering\arraybackslash}m{0.7in}} \hline
\rule{0pt}{3ex} \textbf{Switching Pattern} & $\pmb{T_{on,max}}$ &   $\pmb{T_{on,mid}}$ &  $\pmb{T_{on,min}}$ \\  \hline  
\rule{0pt}{4ex}  \textbf{I} & $|I_{max}| \cdot \frac{T_s}{I_{DC}}$ &   $|I_{mid}| \cdot \frac{T_s}{I_{DC}}$ & $T_s$\\ \hline
\rule{0pt}{4ex}  \textbf{II} & $|I_{max}| \cdot \frac{T_s}{I_{DC}}$ &   $|I_{mid}| \cdot \frac{T_s}{I_{DC}}$ & $|I_{max}| \cdot \frac{T_s}{I_{DC}}$\\ \hline 
\end{tabular} 
      \centering 
       \label{tab:discussion}
         \end{table}

\begin{figure}[hbpt!]
\centering
\includegraphics[scale=0.37]{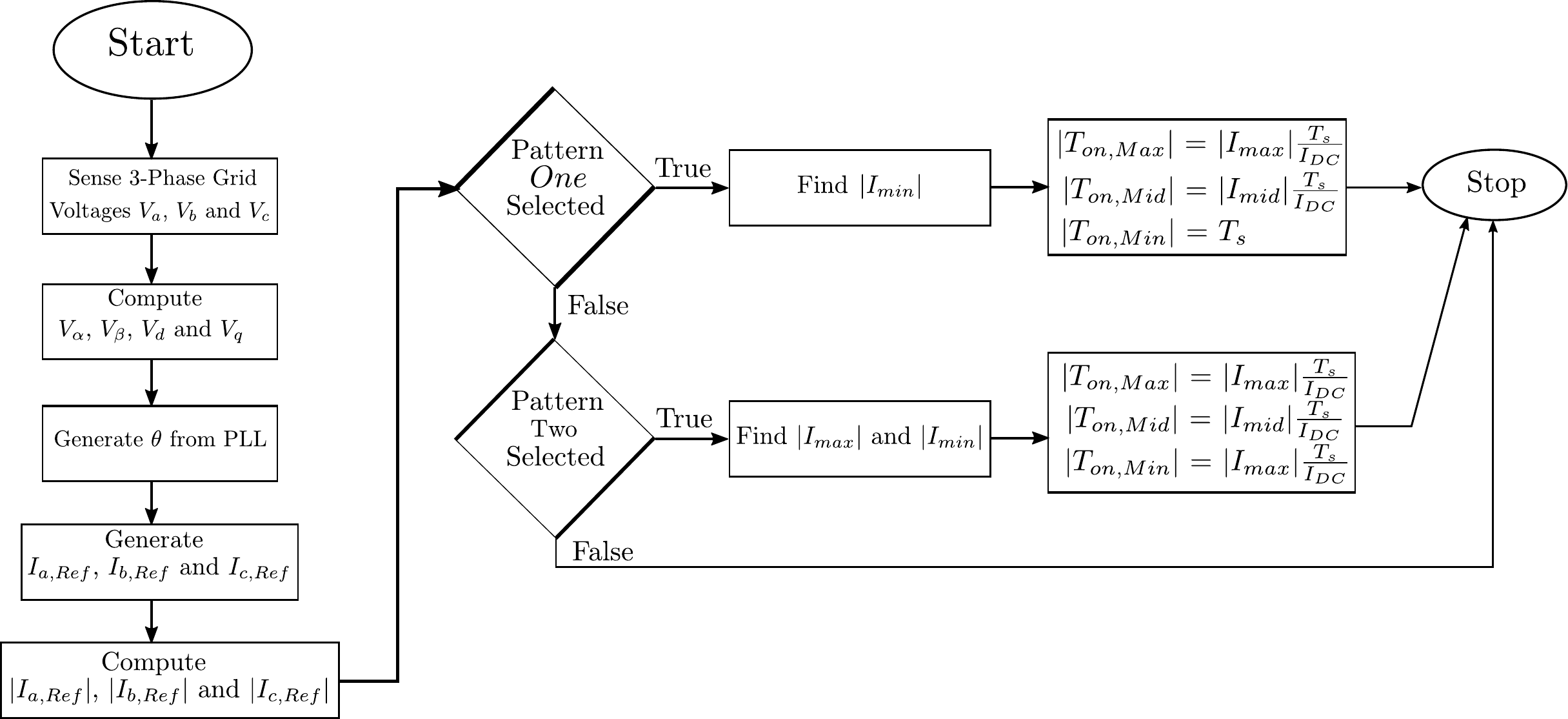}
\caption{Flowchart Depicting the Entire Algorithm of Carrier Generation}
\label{fig:Flowchart}
\end{figure}

\section{Control Algorithm and Hardware Setup}
The control algorithm followed to generate the three carrier waves on a Texas Instruments TMS320F28069 Piccolo$^{TM}$ microcontroller is shown in Fig \ref{fig:Flowchart}. The generic block diagram of the entire hardware setup can be seen in Fig 8, the TPTS converter legs along with output inductors can be seen in Fig \ref{fig:only_tpts_legs}. A 1 kW prototype was built to validate the modulation schemes, the list of components used can be found in Table \ref{tab:components}. A DC electronic load was used in Constant Current (CC) mode which emulated the energy storage element during testing. The carrier wave modulation scheme was tested on the TPTS converter with the DC electronic load drawing 5 A.

\begin{figure}[htbp!] 
\centering
\includegraphics[scale=0.34]{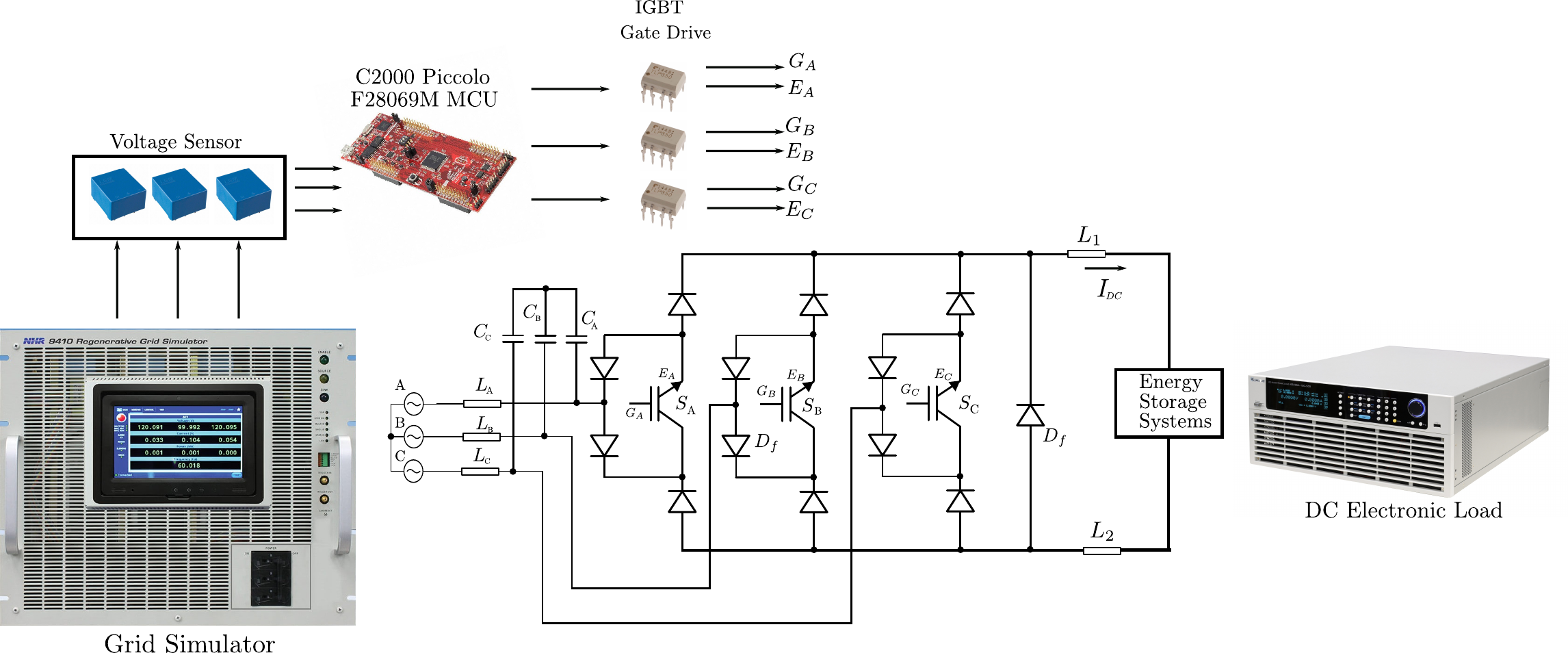}
\caption{Block Diagram of Hardware}
\label{fig:tpts_hardware_block}
\end{figure} 

\begin{figure}[htbp!] 
\centering
\includegraphics[scale=0.25]{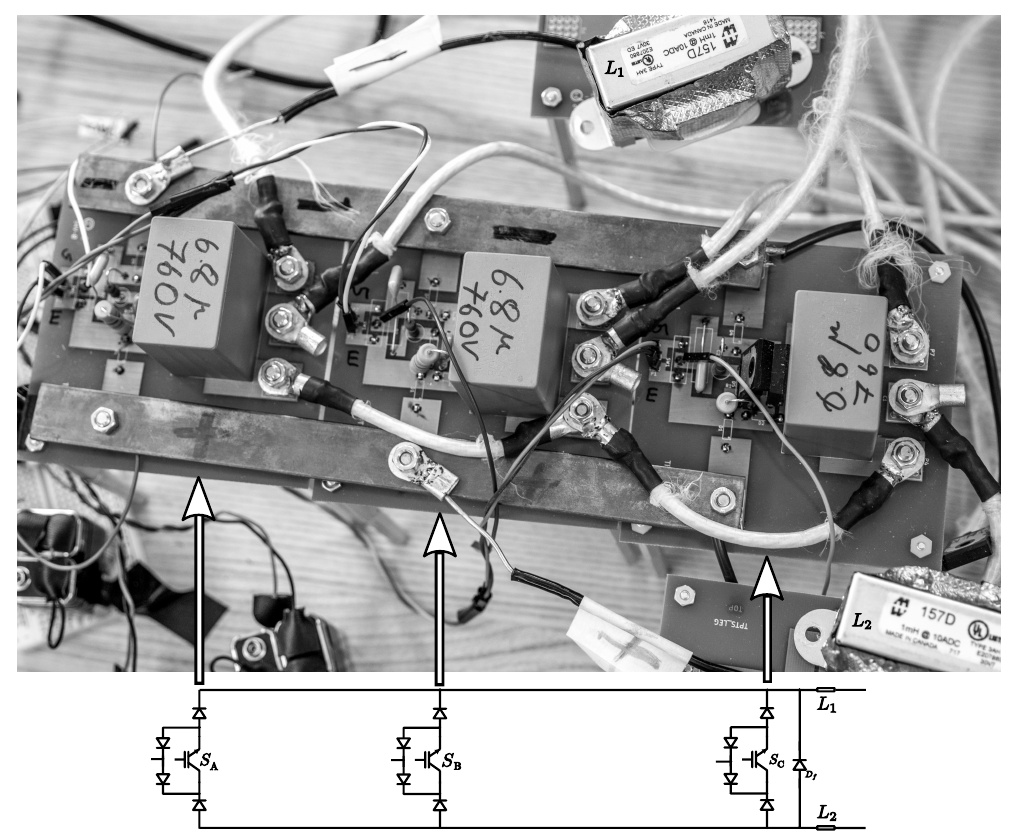}
\caption{TPTS Legs}
\label{fig:only_tpts_legs}
\end{figure} 

\begin{table}[htbp!] 
\caption{{Active and Passive Components Used}}
      \centering
       \begin{tabular}{ >{\centering\arraybackslash}m{1.5in}  >{\centering\arraybackslash}m{1.5in}} \hline
       \textbf{Components} & \textbf{Specification} \\ \hline
 IGBT $S_x$ & IXYH50N120C3, 1200 V, 50 A \\ \hline 
 Diode $D_x$ & STTH75S12, 1200 V, 75 A \\ \hline
 Input Inductor $L_i$ & 230 $\mu$H, 100 m$\Omega$ \\ \hline
 Input Capacitor $C_i$ & 6.8 $\mu$F, 760 $V_{DC}$ \\ \hline 
 Output Inductor $L_o$ & 1 mH, 10 $A_{DC}$ \\ \hline
 Output Capacitor $C_o$ & 150 $\mu$F, 450 $V_{DC}$ \\ \hline
Voltage Sensor &   LV 20-P, 500 $V_{DC}$  \\ \hline
Gate Driver &  TLP350, 2.5 $A_{max}$ \\ \hline
\end{tabular} 
      \centering 
       \label{tab:components}
         \end{table}

  
\section{Results} 
The proposed carrier wave modulation schemes were tested using the TPTS converter at a switching frequency of 18 kHz and a DC load current of 5 A, details of the operating parameters are provided in Table \ref{tab:operating parameters}. The carrier waves generated from switching pattern I and pattern II for a modulation index of 0.5 are shown in Fig 10, Fig 11, Fig 14 and Fig 15 respectively. The phase currents of Phase A and Phase B are shown in Fig 12, Fig 13, Fig 16 and Fig 17. It can be observed that the phase currents track the reference currents. The simulation and experimental results are comparable for a DC load current of 5 A. Comparison of resource utilization between Space vector based modulation scheme and proposed scheme was done on the same hardware and digital controller (TMS320F28069 Piccolo$^{TM}$). From these results it is shown that the proposed carrier-based schemes consumes less resources compared to SV modulation technique for TPTS.


\begin{table}[t] 
\center
 \caption{Operating Parameters}
\begin{tabular}{ >{\centering\arraybackslash}m{1.2in}     >{\centering\arraybackslash}m{1.5in}}
\hline
 \textbf{Parameter}  & \textbf{Specification} \\  \hline  
 $V_{in,line-line,peak}$  & 245 V \\ \hline 
 $V_{dc,0.9}$  & 220 V \\ \hline
 $I_{load,0.9}$  & 5 A \\ \hline
 $f_{sw}$   & 18 kHz \\ \hline
 $P_{out}$ & 1.1 kW \\ \hline 
\end{tabular}
\label{tab:operating parameters}
\end{table}

\begin{table}[!t] 
\caption{Resource Comparison}
      \centering
       \begin{tabular}{  >{\raggedright\arraybackslash}m{1in}  >{\centering\arraybackslash}m{1in}
>{\centering\arraybackslash}m{1in}} \hline
\textbf{Modulation Scheme} & \textbf{Space Vector} & \textbf{Carrier-based}  \\  \hline  
Arithmetic Operators & 8 Multiplication,\newline 2 Addition and 1 Subtraction & 3 Relational Operators and  3 Decision-making constructs  \\ \hline
Trigonometric Functions & Sine and Cosine & None \\ \hline
Look-up Tables & 2 & None \\ \hline
Data Types & Integer and Float & Integer \\ \hline
Operations Involved & Math Library Required  & Only \textit{if-else} statements \\ \hline
Processor Type & Floating-Point Unit & Fixed Point Unit \\ \hline
Memory Usage & 56 KB  & 8 KB \\ \hline
RAM Blocks Used & 7 & 3 \\ \hline
Execution Time & 32 $\mu$s & 8 $\mu$s \\ \hline
\end{tabular} 
      \centering   
      \label{tab:svvscb}    
         \end{table}

\begin{figure}[htbp!] 
\centering
\includegraphics[width=7.5cm,height=4.9cm]{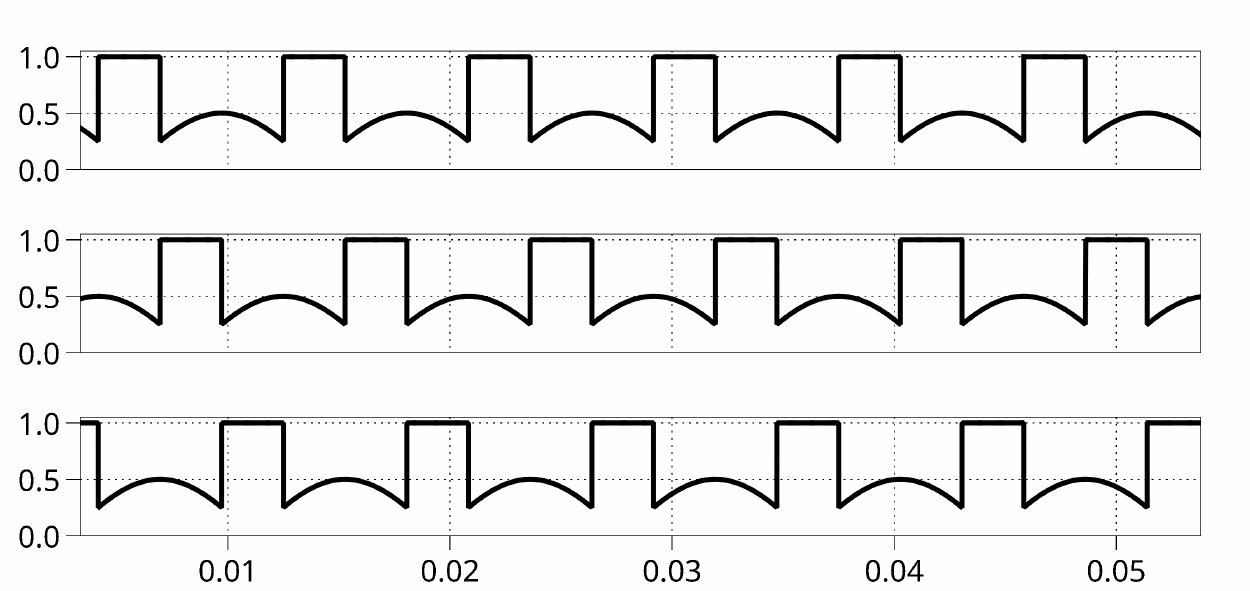}
\caption{Simulation Results at 0.5 modulation index for Switching Pattern I}
\label{fig:tpts_hardware_block}
\end{figure} 

\begin{figure}[htbp!] 
\centering
\includegraphics[width=7.5cm,height=4.9cm]{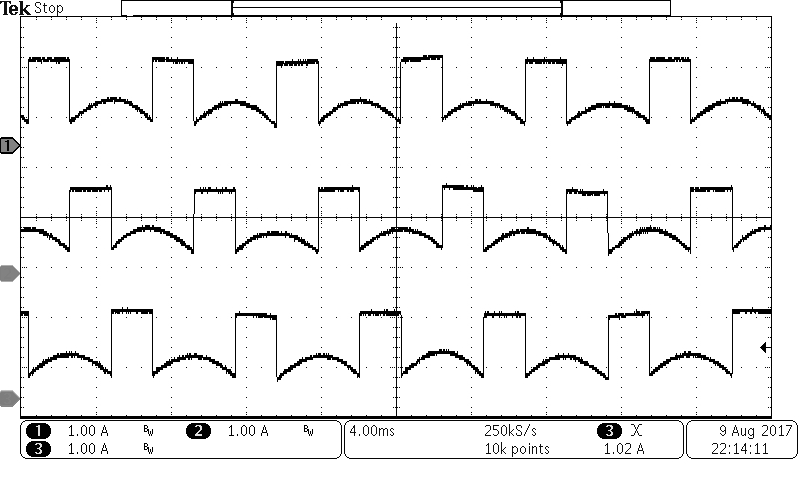}
\caption{Experimental Results at 0.5 modulation index for Switching Pattern I}
\label{fig:tpts_hardware_block}
\end{figure}

\begin{figure}[htbp!] 
\centering
\includegraphics[width=7cm,height=4.3cm]{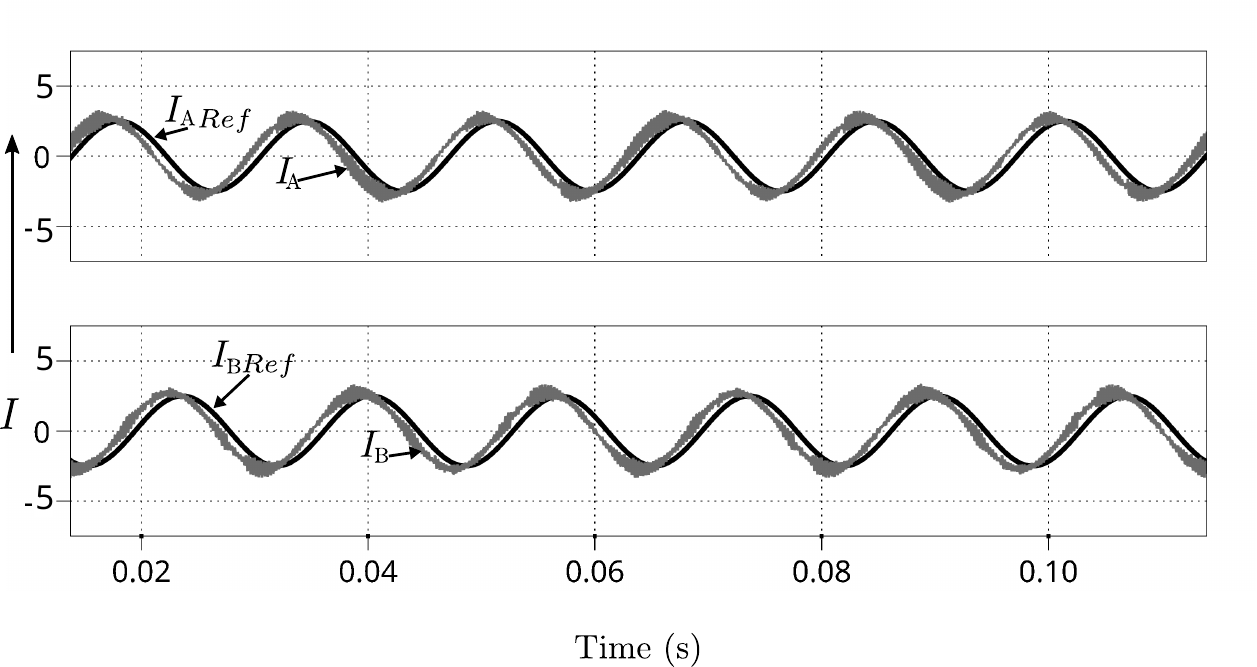}
\caption{Simulation Results depicting phase currents at 0.5 modulation index for Switching Pattern I}
\label{fig:tpts_hardware_block}
\end{figure} 

\begin{figure}[htbp!] 
\centering
\includegraphics[scale=0.3]{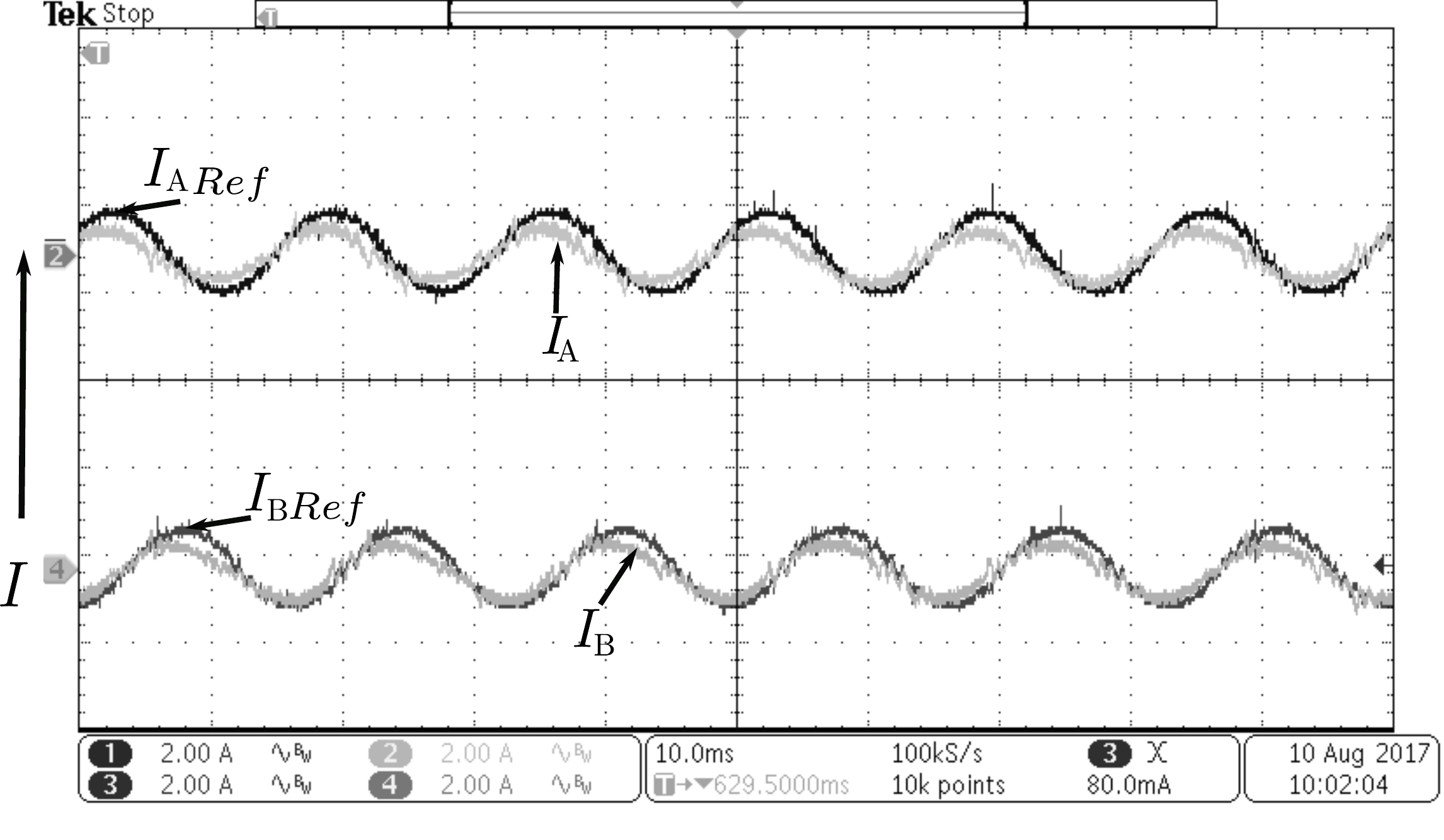}
\caption{Experimental Results depicting phase currents at 0.5 modulation index for Switching Pattern I  (current scale: 5 A/div, time scale: 10 ms/div)}
\label{fig:tpts_hardware_block}
\end{figure} 


\begin{figure}[htbp!] 
\centering
\includegraphics[width=7cm,height=4.5cm]{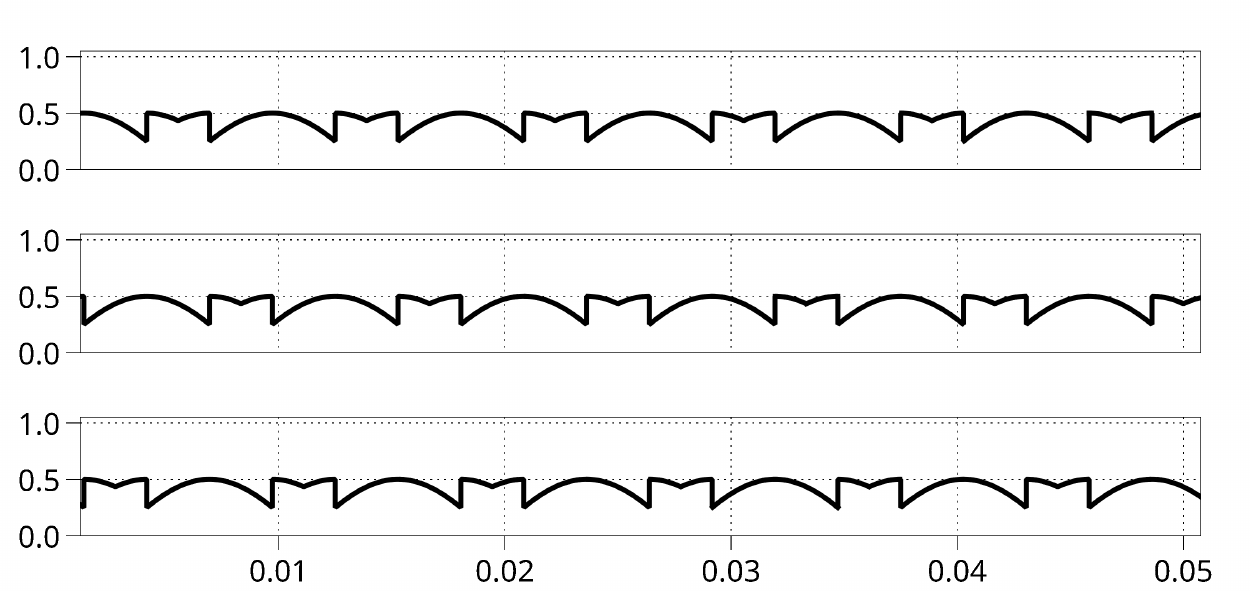}
\caption{Simulation Results at 0.5 modulation index for Switching Pattern II}
\label{fig:tpts_hardware_block}
\end{figure} 

\begin{figure}[htbp!] 
\centering
\includegraphics[width=7cm,height=4.5cm]{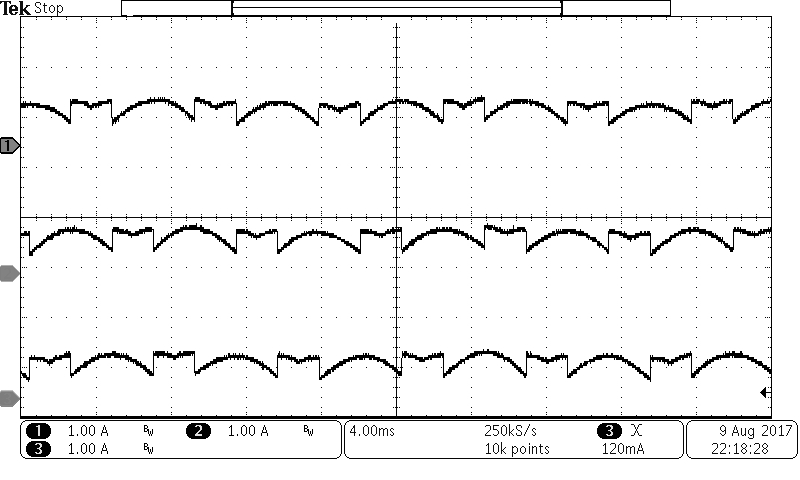}
\caption{Experimental Results at 0.5 modulation index for Switching Pattern II}
\label{fig:tpts_hardware_block}
\end{figure}

\begin{figure}[htbp!] 
\centering
\includegraphics[width=7cm,height=4.3cm]{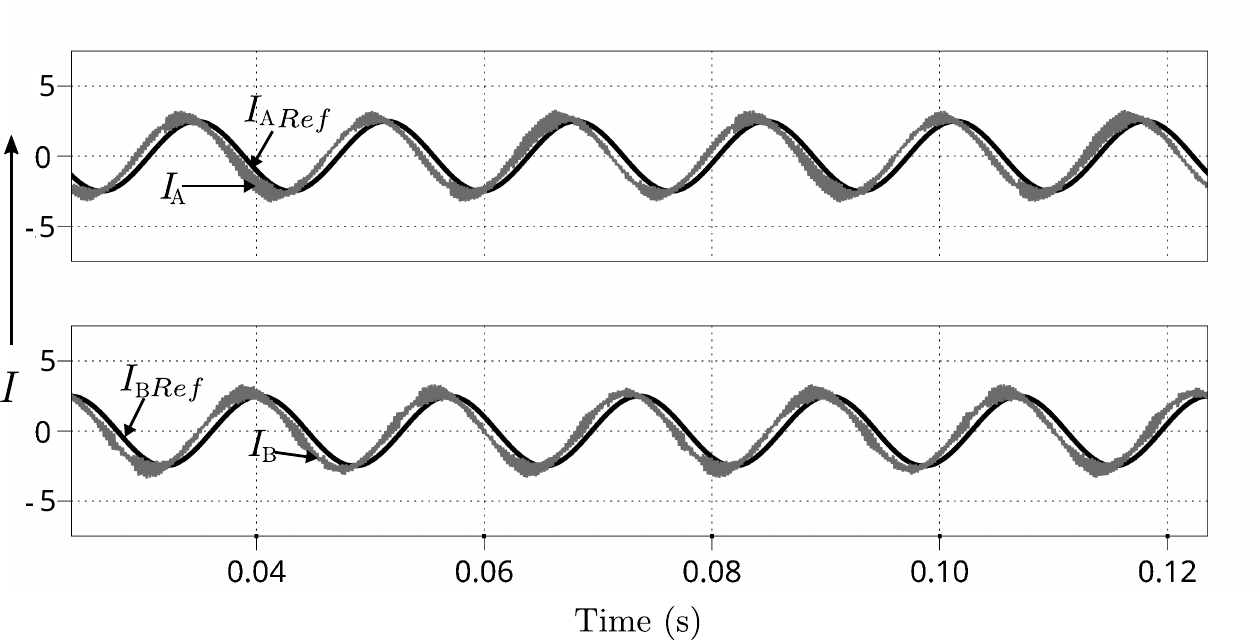}
\caption{Simulation Results depicting phase currents at 0.5 modulation index for Switching Pattern II}
\label{fig:tpts_hardware_block}
\end{figure} 

\begin{figure}[htbp!] 
\centering
\includegraphics[scale=0.3]{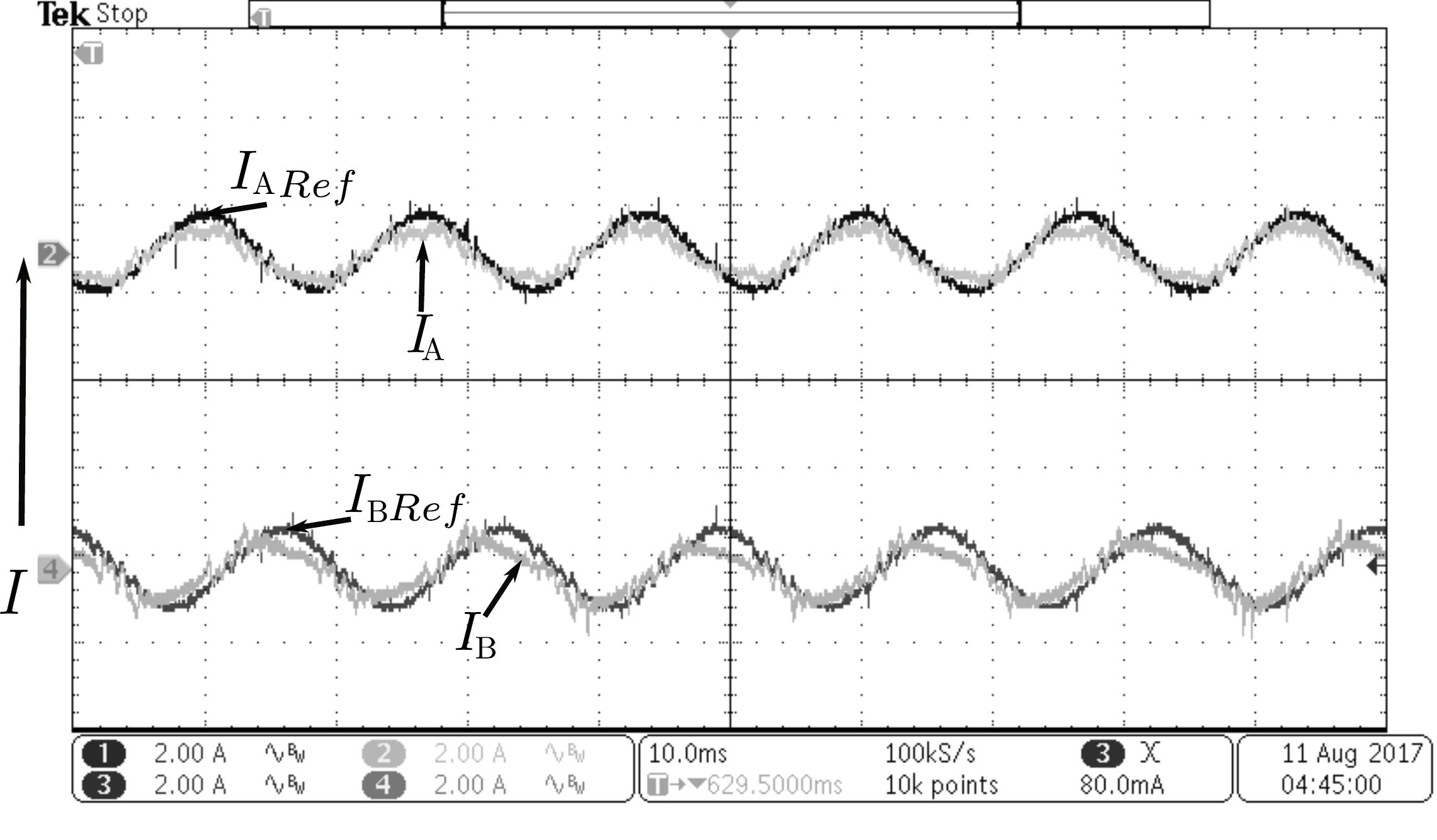}
\caption{Experimental Results depicting phase currents at 0.5 modulation index for Switching Pattern II  (current scale: 5 A/div, time scale: 10 ms/div)}
\label{fig:tpts_hardware_block}
\end{figure}

\section{Conclusions}
Selecting a reliable and efficient power converter to charge EV batteries is a challenging task. A TPTS rectifier is an ideal candidate for DC fast charging applications, due to its efficiency at peak power rating. This paper introduces two simplified carrier based modulation schemes which greatly reduce the burden of implementation as they involve minimum  mathematical computations. The proposed modulation scheme is compared with the conventional SV modulation technique based on resource utilization on a digital controller. The carrier-based technique consumes very less resources and hence, is preferred over the conventional SV modulation scheme. This paper also provides detailed procedure to generate each carrier wave from the  available  three phase grid power supply. The performance comparison of the proposed modulation schemes are validated using simulation and experimental results. 
  
%
\IEEEpeerreviewmaketitle



\begin{thebibliography}{1}

\bibitem{evs} D. A. Giménez-Gaydou, A. S. N. Ribeiro, J. Gutiérrez, and A. P. Antunes, ``Optimal location of battery electric vehicle charging stations in urban areas: A new approach,'' \textit{International Journal of Sustainable Transportation,} vol. 10, no. 5, pp. 393-405, May 2016.

\bibitem{ptk} M. Yilmaz and P. Krein, ``Review of battery charger topologies, charging power levels, infrastructure for plug-in electric and hybrid vehicles,'' \textit{IEEE Transactions on Power Electronics}., vol. 28, no. 5, pp. 2151-2169, May 2013.


\bibitem{t1} V. Monteiro, J. C. Ferreira, A. A. Melendez, C. Couto, and J. L. Afonso, ``Experimental Validation of a Novel Architecture Based on a Dual-Stage Converter for Off-Board Fast Battery Chargers of Electric Vehicles,'' \textit{IEEE Transactions on Vehicular Technology,} vol. PP, no. 99, pp. 1-8, 2017.

\bibitem{jan}  J. Channegowda, Vamsi Krishna Pathipati and Sheldon S. Williamson ``Comprehensive Review and Comparison of DC Fast Charging Converter Topologies: Improving Electric Vehicle Plug-to-Wheels Efficiency,'' in \textit{Proceedings IEEE International Symposium On Industrial Electronics}, pp. 1-6, Buzios, Rio
de Janeiro, Brazil, June. 3-5, 2015.



\bibitem{kodes} T Nussbaumer, M Baumann, and J W. Kolar, `` Comprehensive Design of a Three-Phase Three-Switch Buck-Type PWM Rectifier,'' \textit{IEEE Transactions on Power Electronics}., vol. 22, no. 2, pp. 551 - 562, March 2007.

\bibitem{newwide}M. Baumann, U. Drofenik, and J. W. Kolar, ``New Wide Input Voltage Range Three-Phase Unity Power Factor Rectifier Formed by Integration of a Three-Switch Buck-Derived Front-End and a DC/DC Boost Converter Output Stage,'' in \textit{Proc, IEEE International Telecommunications Energy Conference}, Phoenix, Arizona, U.S.A., pp. 461-470, Sept. 14 - 18, 2000. 


\bibitem{k1} M. Baumann and J.W. Kolar, ``Comparative evaluation of modulation methods for a three-phase/switch buck power factor corrector concerning the input capacitor voltage ripple,'' in \textit{Proc. 32nd IEEE Power Electronics Specialists Conference}, Vancouver, Canada,pp. 1327-1332, June 17-21 2001.

\bibitem{car1}  Beomseok Chae, and Yongsug Suh ``Improved operating range for three-phase three-switch buck-type rectifier using carrier based PWM ,'' \textit{Proc. IEEE 8th International Power Electronics and Motion Control Conference (IPEMC-ECCE Asia)}, May 2016, pp. 2302-2309.



\bibitem{emil}O Dordevic, M Jones and E Levi, `` A Comparison of Carrier-Based and Space Vector
PWM Techniques for Three-Level Five-Phase Voltage Source Inverters,''  \textit{IEEE Transactions on Industrial Informatics}, vol 9, no. 2, pp. 609 - 619, May 2013.
 
\bibitem{naj} N Azeez, A Dey, K. Mathew, J Mathew, K. Gopakumar,  and M P. Kazmierkowski ``A Medium-Voltage Inverter-Fed IM Drive Using Multilevel 12-Sided Polygonal Vectors, With Nearly Constant Switching Frequency Current Hysteresis Controller,'' \textit{IEEE Transactions on Industrial Electronics}, vol 61, no. 4, pp. 1700 - 1709, April 2014.


\bibitem{janapec}J. Channegowda, N. A. Azeez, and S. S. Williamson, ``Simplified Carrier-Based Modulation Scheme for Three-Phase Three-Switch Rectifier for DC Fast Charging Applications,'' in \textit{Proc. Applied Power Electronics Conf. and Expo.,} Tampa, FL, March 2017 pp. 3501 - 3506. 



 

 

\end{thebibliography}
\end{document}